\documentclass[12pt,preprint]{aastex}
\begin{document}
\newcommand{\up}[1]{\ifmmode^{\rm #1}\else$^{\rm #1}$\fi}
\newcommand{\zdot}{\makebox[0pt][l]{.}}
\newcommand{\upd}{\up{d}}
\newcommand{\uph}{\up{h}}
\newcommand{\upm}{\up{m}}
\newcommand{\ups}{\up{s}}
\newcommand{\arcd}{\ifmmode^{\circ}\else$^{\circ}$\fi}
\newcommand{\arcm}{\ifmmode{'}\else$'$\fi}
\newcommand{\arcs}{\ifmmode{''}\else$''$\fi}

\title{The Araucaria Project. Determination of the
LMC Distance from Late-Type Eclipsing Binary Systems: 
I. OGLE-051019.64-685812.3
\footnote{Based on observations obtained with the ESO NTT for 
Programmes 074.D-0318(B) and 074.D-0505(B), and with the Magellan Clay telescope
at Las Campanas Observatory}
}
\author{Grzegorz Pietrzy{\'n}ski}
\affil{Universidad de Concepci{\'o}n, Departamento de Astronomia,
Casilla 160-C, Concepci{\'o}n, Chile}
\affil{Warsaw University Observatory, Al. Ujazdowskie 4, 00-478, Warsaw, Poland}
\authoremail{pietrzyn@astrouw.edu.pl}
\author{Ian B. Thompson}
\affil{Carnegie Observatories, 813 Santa Barbara Street, Pasadena, CA,
911101-1292}
\author{Dariusz Graczyk}
\affil{Universidad de Concepci{\'o}n, Departamento de Astronomia,
Casilla 160-C, Concepci{\'o}n, Chile}
\authoremail{Darek.Graczyk@astri.uni.torun.pl}
\author{Wolfgang Gieren}
\affil{Universidad de Concepci{\'o}n, Departamento de Astronomia,
Casilla 160-C, Concepci{\'o}n, Chile}
\authoremail{wgieren@astro-udec.cl}
\author{Andrzej  Udalski}
\affil{Warsaw University Observatory, Al. Ujazdowskie 4, 00-478, Warsaw,
Poland}
\authoremail{udalski@astrouw.edu.pl}
\author{Olaf Szewczyk}
\affil{Universidad de Concepci{\'o}n, Departamento de Astronomia,
Casilla 160-C, Concepci{\'o}n, Chile}
\authoremail{szewczyk@astro-udec.cl}
\author{Dante Minniti}
\affil{Departamento de Astronomia y Astrofisica, Pontificia Universidad
Cat{\'o}lica de Chile, Casilla 306, Santiago 22, Chile}
\authoremail{dante@astro.puc.cl}
\author{Zbigniew  Ko{\l}aczkowski}
\affil{Universidad de Concepci{\'o}n, Departamento de Astronomia,
Casilla 160-C, Concepci{\'o}n, Chile}
\authoremail{zibi@astro-udec.cl}
\author{Fabio Bresolin}
\affil{Institute for Astronomy, University of Hawaii at Manoa, 2680 Woodlawn 
Drive,  Honolulu HI 96822, USA}
\authoremail{bresolin@ifa.hawaii.edu}
\author{Rolf-Peter Kudritzki}
\affil{Institute for Astronomy, University of Hawaii at Manoa, 2680 Woodlawn 
Drive, Honolulu HI 96822, USA}
\authoremail{kud@ifa.hawaii.edu}

\begin{abstract}
We have analyzed the double-lined eclipsing binary system OGLE-051019.64-685812.3 
in the LMC which
consists of two G4 giant components with very similar effective temperatures. A
detailed analysis of the OGLE I-band light curve of the system, radial velocity curves
for both components derived from high-resolution spectra, and near-infrared magnitudes
of the binary system measured outside the eclipses has allowed us to obtain
an accurate orbit solution for this eclipsing binary, and its fundamental physical
parameters. Using a surface brightness-(V-K) color relation for giant stars 
we have calculated the distance to the system and obtain a true distance modulus
of 18.50 mag, with an estimated total uncertainty of $\pm$ 3\%. More similar eclipsing
binary systems in the LMC which we have discovered and for which we are currently
obtaining the relevant data will allow us to better check on the systematics of
the method and eventually provide a distance determination to the LMC accurate
to 1 percent,
so much needed for the calibration of the distance scale.

\end{abstract}


\section{Introduction}
An accurate calibration of the extragalactic distance scale is one of the
most important and challenging tasks of modern astronomy. In spite of 
substantial improvements achieved over the years in this field 
there are still important systematic errors associated with both the primary Cepheid and
other secondary methods which do not yet allow to obtain extragalactic
distances, and thus the Hubble constant with the high accuracy 
needed for cosmological applications. The
major uncertainties in the use of Cepheid variables as standard candles continue to be
our lack of a detailed understanding of the effect of metallicity on  
the Period-Luminosity (PL) relation (e.g. Romaniello et al. 2008; Groenewegen 2008:
Gieren et al. 2005; Storm et al. 2004) and, perhaps
most importantly,  the distance to the LMC. The LMC  provides the fiducial
Cepheid PL relation, extremely well established from the OGLE microlensing survey in 
optical bands (Udalski et al. 1999a; Soszynski et al. 2008), and by the work of Persson et al.
(2004) in the near-IR JHK bands. The distance to the LMC is likely to constitute the largest source of uncertainty
in the construction of the extragalactic distance ladder with the Cepheid method (Freedman et
al. 2001). In our ongoing Araucaria project, we are addressing these
issues in an effort to reduce  the current systematic uncertainties in both
the metallicity effect on the Cepheid PL relation and  
the distance to the LMC.

Detached eclipsing doubled-lined spectroscopic binaries offer a unique
opportunity to directly measure  accurate stellar parameters
like mass, luminosity,  radius (Andersen 1991), and  distance 
(Paczynski 1997). It has been argued (Paczynski 1997; Paczynski 2000) 
that with current observational facilities, eclipsing binaries carry the potential
to yield the most direct (one step) and accurate (about 2-3\%)
distance to the LMC. For an extensive historical
review of this technique the reader is referred to the paper of
Kruszewski and Semeniuk (1999), while the method itself is very well 
described by Lacy (1977), and  Paczynski (1997). 
Briefly, using high-quality radial velocity and photometric observations,
standard fitting routines (e.g. Wilson and Devinney 1971) provide very 
accurate masses, sizes, and surface brightness ratios for the components
of a double-lined eclipsing binary (e.g. Andersen 1991). The distance 
to the system follows from the dimensions determined this way, plus the
absolute surface brightness, which can be inferred from the observed stellar
colors from a precise empirical surface brightness-color calibration. Such relations,
for different colors, are well established  for stars
with spectral types later than A5 from accurate measurements 
of stellar angular diameters by interferometry (di Benedetto 1998;
Kervella et al. 2004, Groenewegen 2004, di Benedetto 2005). 

This powerful technique has been used by Thompson et al. (2001) 
to determine the distance to $\omega$ Centauri using observations of
the cluster eclipsing binary OGLEGC17. 
Unfortunately, late-type main sequence binary stars located in the LMC
are too faint to secure high resolution spectra for accurate radial 
velocity measurements even with the largest current telescopes. Therefore 
the few attempts made to use eclipsing binaries to determine the distance 
to the LMC (e.g. Guinan et al. 1998) have used relatively bright early-type 
systems, for which however an accurate empirical 
surface brightness -- color relation is currently not available. As a consequence, 
theoretical models have to be employed in this work, this has prevented the full
realization of the potential offered by  eclipsing binaries for an 
accurate measurement of the distance to the LMC.

In order to improve on this situation
our group has performed a careful analysis of the optical (BVI) light curves of the LMC
eclipsing binary stars catalogued by the OGLE consortium (Wyrzykowski et al.
2003). We have used the Wilson-Devinney (WD) code to identify eight 
long-period eclipsing systems composed of {\it two late-type giant components}. 
These systems offer for the first time the opportunity to take full advantage of
the eclipsing binary method and measure a truly accurate 
distance to the LMC, using the empirical  surface brightness -- color relation
which is very well established for late-type stars. 
During the past two years we have been obtaining near-infrared 
and high-resolution spectroscopic observations for all these systems. 
In this paper, we present the first distance measurement 
to the LMC based on our to-date best observed late-type eclipsing binary system,
OGLE-051019.64-685812.3.

\section{Observations, Data Reduction and Calibration}

\subsection{Optical Photometry}
The optical photometry of our binary system was obtained with the Warsaw
1.3m telescope at Las Campanas Observatory in the course of the
second and third phases of the OGLE project (Wyrzykowski et al. 1999). 
A total of 780  $I$-band  epochs spanning a period of 4124 days were
secured. In order to obtain accurate colors, we also collected  75 $V$-band
band images. The data were reduced with the 
image-subtraction technique (Udalski 2003, Wozniak 2000). 
The instrumental data were very carefully
calibrated onto the standard system using observations of several 
Landolt fields over many dozens of photometric nights. The estimated 
zero point errors are about 0.01 mag in all bands (Udalski et al. 1999b).
For more details about the instrumental system, observing, reduction and
calibration procedures adopted in the course of the OGLE project 
the reader is referred to the references cited above.

\subsection{Near-Infrared Photometry}
The near-infrared data presented in this paper were collected with the
ESO NTT telescope on La Silla, equipped with the
SOFI infrared camera. We used the Large Field setup with a field
of view of 4.9 x 4.9 arcmin at a scale of 0.288 arcsec/pixel.
The gain and readout noise were 5.4~e/ADU and 0.4~e, respectively.
The data were obtained in two observational programs,
074.D-0318(B), 074.D-0505(B) (PI: Pietrzy{\'n}ski) as part
of the Araucaria Project.
Single deep $J$-band and $K_s$-band observations of  our target field
were obtained under excellent seeing conditions during five different
nights.  On these nights, we also observed a large number 
(8-12) of photometric standard stars from the UKIRT system (Hawarden et al.
2001) at a variety of airmasses and spanning a broad range in colors.

To account for the frequent sky level variations in the infrared
spectral region, especially in the $K_s$-band, the observations were performed with a
dithering technique. For the $K_s$ and $J$-band observations we  averaged over 10 consecutive 10 second
integrations (DITs)  at any given pointing before moving the telescope to a randomly
selected different position within a 25$\times$25 arcsec square.
Between 15 and 25 such dithering positions were obtained through the $K_s$-band. 
In the case of the $J$-band filter this number varied
from 11 to 15. 

The reductions were performed in a similar manner to those
described in Pietrzy{\'n}ski and Gieren (2002). The sky was subtracted
from the images with a two-step process implying masking
of the stars with the xdimsum IRAF package. Then the individual
images for each field and filter were flatfielded and stacked into a
final composite image.
The PSF photometry was carried out with the DAOPHOT and ALLSTAR
programs.
About 20-30 relatively bright and isolated stars were selected visually
and the first PSF model was derived from them. 
Following Pietrzy{\'n}ski, Gieren and Udalski (2002),  we then iteratively
improved the PSF model by subtracting all stars from their neighbourhood and
re-calculating the PSF model. After three such iterations no further improvement was noted,
and the corresponding PSF model was  adopted.
                        
In order to convert our PSF photometry to the aperture system, aperture
corrections were derived for each frame. This was done by using the previously
identified candidates for PSF calculations after removing all nearby stars that could
contaminate our photometry. The median from the aperture corrections derived for these
stars was adopted as a aperture correction for a given frame. Typically,
the rms scatter in the aperture corrections derived in this way was
better than 0.01~mag.

Aperture photometry on the standard stars was performed by choosing an
aperture radius of 16 pixels. Four of our five nights were photometric 
and the tranformations onto the UKIRT system were performed
independently for each of them. The total error of the
zero points of the $K_s$ and $J$ magnitudes is estimated
to be about 0.02~mag for any given night.

\subsection{High Resolution Spectroscopy}
Echelle spectra were collected with the Magellan 
Clay 6.5 m telescope equipped with the MIKE  spectrograph.
A single 5 x 0.7 arcsec slit was used during the observations 
giving  a  resolution of  about 40 000. A total of 11 spectra
with net exposure times of 1 hour each were collected. In order to allow for
better cosmic ray removal each observation was divided into two 
consecutive 0.5 hour exposures. The spectra were reduced with the 
pipeline software developed by Daniel Kelson, following Kelson (2003).  
The resulting S/N ratio  was about 10 at a wavelenght of 4500~\AA.
Radial velocities were measured with the TODCOR package (Mazeh and  
Zucker 1994) in the wavelength regions 4125 - 4320, 4350 - 4600, 
and 4600 - 4850 \AA   using templates taken from the Coelho et al. (2005) 
synthetic library.  The final radial velocity for
a given spectrum was adopted as
the mean from the measurements obtained from the different wavelength ranges.
The individual RV measurements are listed in Table 1.

\section{Spectroscopic and Photometric Solutions}

The following final
ephemeris was derived from the photometric data for our binary system
based on the AoV technique (Schwarzenberg-Czerny 1996): \\
                                                                        
% $ {\rm P}  =  214.370 \pm  0.008   days  \hspace*{2cm}  {\rm T}_{\rm 0} = 2450390.127 \pm 0.009 $ \\

$ {\rm P}  =  214.370 \pm  0.008 \hspace*{0.1cm} days  \hspace*{1cm}  {\rm T}_{\rm 0} = 2450498.0 \pm 0.1 $ days\\

We adopt the photometric ephemeris and derive the systemic velocity,
the velocity amplitudes, eccentricity, periastron passage, and the mass ratio
from a least squares fit to the velocity data (see Fig. 1 and  Table 3). 
The measured systemic velocity confirms that our binary star belongs to the LMC. 

The mean (constant) $V$ (16.655 mag), $I$ (15.685 mag), and $K$ (14.437 mag) band magnitudes of the system outside
eclipse were derived as the average 
of all of the  individual out-of-eclipse observations obtained in the corresponding band.
A reddening of E(B-V) = 0.146 $\pm$ 0.02 mag was 
adopted for the binary from the OGLE extinction maps (Udalski et al. 1999b).

Figure 2 shows that the observed minima have approximately the same depths, indicating that the
effective temperatures of the components must be similar. 
Therefore we can assume that the difference in brightness is caused by
the difference in the temperatures and
taking the effective temperature of the whole system as derived
from the observed colors, and using  several independent
empirical  calibrations (see Table 2), calculate 
the effective temperatures of the individual 
components.
Effective temperatures of 
$T_{1} = 5300 \pm 100  K$ and $T_{2} = 5450 \pm 100 K$, were 
obtained in this way for the primary and secondary components,
respectively. 
Next, the limb darkening coefficients corresponding to  the logarithmic formula
were taken from the tables of van Hamme (1993) by assuming a metallicity of 
[Fe/H] = -0.5 dex (e.g. the typical metallicity  for the red giant stars in 
the LMC; Olszewski et al. 1991, Cole et al. 2000), and adopting the 
gravities and effective temperatures obtained from the preliminary WD 
solution. We adopted bolometric albedo and gravity brightening coefficients 
of A = 0.5 and g = 0.32, respectively, values appropriate 
for stars with convective envelopes. The mass ratio 
was fixed to the spectroscopic value of 0.9695, and we assumed that
there is no light contribution from a third body. 

The shape of the I band light curve clearly suggests that the 
binary is well detached, so mode 2 of the WD code was chosen.
To obtain consistency with the spectroscopic solution a grid of photometric
solutions was calculated for different values of the eccentricity 
in the range of values consistent with the spectroscopic data 
(e.g.  0.034 - 0.042). 
The following parameters were 
adjusted: the temperature of the secondary component ($T_{2}$), the dimensionless 
potentials $\Omega_{1}, \Omega_{2}$, luminosity ($ L_{1}$), inclination (i), 
the periastron pasage ($\omega$), eccentricity (e) and the phase displacement ($\phi$).
The best solution (e.g. with the smallest $\chi^2$) was obtained for 
e = 0.0395, and the corresponding
parameters listed in Table 3. 
In order to assign realistic errors to
these parameters, we computed several grids of models for potentially degenerate
parameters and analyzed the corresponding space parameters. 
The errors derived in this way are also given in Table 3.
We also computed a series of models assuming a light contribution 
from a third body of 2, 4 and  6 \%.  In each case we found that adding this 
third light source degraded the quality of our solutions  significantly, 
which confirms our preliminary  assumption that there is no third light 
source in this system.

We conclude that  OGLE-051019.64-685812.3 is a completely 
detached, slightly eccentric system composed of two G4III giants, and therefore 
is ideal for an accurate distance determination.

\section{Distance Determination}

OGLE-051019.64-685812.3 offers us an outstanding opportunity to derive a very accurate  
distance to the LMC using an empirical surface brightness -- color
relation (di Benedetto 1998, Groenewegen 2004,  Kervella et al. 2004, 
di Benedetto 2005). 
We would like to note here, that another calibration given by van Belle (1999) 
is not accurate enough for our purpose (e.g. about 10 \% only, due to the lower
accuracy of their measured angular diameters). 

Given the long period of the system it was not possible to obtain eclipse
light curves in all passbands, and as a result we computed
the magnitudes and colors of the individual components 
by deriving the brightness ratios in different bands from an 
extrapolation of the $I$-band solution (e.g. taking into account 
the information about the temperatures of both components). 

Since all available empirical surface brightness-color relations are based on magnitudes 
calibrated onto the Johnson system, we transformed our photometry to this
system using the equations given by Carpenter et al. (2001), and Bessell and Brett (1988).
Adopting the reddening law of Schlegel et al. (1998) and a reddening of E(B-V) = 0.146 mag
as derived from the OGLE reddening maps, the reddening-free magnitudes and colors 
of the individual components were calculated. These are listed 
in Table 3.

The distance in parsecs follows then directly from the Lacy (1977) equation:

$ d(pc) = 1.337 \times 10^{-5}r(km)/ \varphi(mas)$

The linear diameter (r) comes from the analysis of the system while 
the angular diameter ($\varphi$) is derived from the surface brightness-color 
relations (e.g.  $ m_{0} = S - 5 log(\varphi)$, where S is 
the surface brightness in a given band, and $ m_{0}$ is the unreddened magnitude 
of a given star in this band). 

Calculating the respective surface brightnesses of the components of
OGLE-051019.64-685812.3 from their (V-K) colors and using the calibration of
di Benedetto
(2005) obtained for a mixed sample of giant and dwarf stars, we obtain distances of 
(50.4 $\pm$ 1.3) kpc for the primary,
and (50.0 $\pm$ 1.4) kpc for the secondary component, corresponding to distance moduli 
of (18.51 $\pm$ 0.06) mag and (18.49 $\pm$ 0.06) mag, respectively.
Very similar results (see Table 2) are obtained
using the (V-K) colors and the calibrations of di Benedetto (1998), Groenewegen  (2004),
Kervella (2004).

It has been shown by Di Benedetto (1998, 2005) that the surface brightness -- color relations
for late-type dwarf and giant stars are consistent with each other 
at the level of 1 \% . To demonstrate the very low sensitivity of our
derived distance value on the adopted surface brightness - color relation, we present
in Table 2 the distances of the two components
of OGLE-051019.64-685812.3 which we obtain using the surface brightness - color relations based on
dwarfs (di Benedetto 1998, Groenewegen 2004, Kervella et al. 2004),
giants (di Benedetto 1998, Groenewegen 2004), and a mixed sample of both types of stars
(di Benedetto 1998, di Benedetto 2005). The distance results agree extremely well.
We adopt for the distance to OGLE-051019.64-685812.3 the value obtained from the
most recent di Benedetto (2005) surface brightness-color relation, (18.50 $\pm$ 0.06) mag,
because this particular relation
is based on a large  number of very carefully selected calibrating angular diameter stars.

\section{Discussion}

In the previous paragraph we decided to use the $V-K$ color to derive the surface brightnesses
and distances because the corresponding surface brightness -- color relation has the smallest scatter
(e.g. Di Benedetto 1998, Kervella et al. 2004, Groenewegen 2004, Di Benedetto 2005).  
However, the use of other colors, for example $V-I$, leads to very consistent distance
results. 

The metallicity dependence of the surface brightness is very weak 
(Thompson et al. 2001, Di Benedetto 1998, Groenewegen 2004). The corrections 
for metallicity for our system are as small as 0.007 mag 
for the $K$-band filter. Therefore we decided to neglect them and 
assume an additional error of 0.007 mag in the total error budget. 

In order to calulate the total uncertainty on our distance determination, the
uncertainties on 
the following quantities were taken into account: $ K_{1}$  and $ K_{2}$
(0.5~\%), (the absolute dimension is known to 0.5 ~\%), 
inclination (0.2 \%),
relative radii (1.2 ~\%), the zero point of the optical (0.6~\%) and near infrared 
(0.8~\%) photometries, reddening (0.8~\%), and the error associated with the calibration 
of the surface brightness - color relation (2.0 ~\%).
Adding the contributions of these errors on the distance determination
to OGLE-051019.64-685812.3 quadratically we obtain a total error of 3~\%. We note
that any correction for the tilt of the LMC bar with respect to the line of sight
is expected to be very  small ($<$~0.01 mag) for our target, due to its position
very close to the center of the LMC bar.

Finally, we note that our distance determination is only very weakly dependent
on the assumed reddening. This is because the surface brightness - color relation we used
is almost parallel to the reddening line. Indeed, if we change the adopted reddening 
by as much as 0.06 mag (e.g. 3 $\sigma$) the distance will change by 1 \% only.
If we change the reddening law  by any reasonable amount 
(e.g. $R_{V}$ = 2.7) the effect on the distance 
will be  0.3 \%, and therefore negligible.

Since our system is located very close to the barycenter of the LMC, 
we can assume that its distance represents very closely the distance 
to the LMC. However, we cannot exclude that the system is located 
slightly in front or behind the LMC barycenter. We will investigate 
such a possible depth effect once we have analyzed more late-type
eclipsing binaries for which we are currently acquiring the necessary
data.  

\section{Conclusions}
We have studied the double-lined late-type LMC eclipsing binary system OGLE-051019.64-685812.3
which we discovered in the database of the OGLE-II Project.
From a detailed analysis of its I-band light curve, the radial velocity curves
for the two giant components of the system, and near-infrared photometry outside the eclipses
we have derived the distance of the system from the
$V-K$ surface brightness-color relation for giant stars given by Di Benedetto (2005). 
We obtain a true distance modulus of 18.50 mag, with a total estimated uncertainty 
of 3\%, or 0.06~mag. This result constitutes a significant improvement on the distance 
measurements using  observations of early-type 
eclipsing binaries in the LMC (e.g. Guinan et al. 1998, Ribas et al. 2002).

Our derived distance for OGLE-051019.64-685812.3  agrees very well with most recent
determinations  of the LMC distance from different methods (e.g. Schaefer 2008, Benedict et al. 2007, 
Fouqu{\'e} et al. 2007, Guinan et al. 2004).

This is the first of a number of similar eclipsing binary systems we have discovered
in the LMC, whose study in forthcoming papers will allow us to reduce the uncertainty on the current distance
determination to the LMC from OGLE-051019.64-685812.3. The
long orbital periods and faint magnitudes of these systems make the collection of
the required photometric and spectroscopic data difficult, but our present results
demonstrate the high potential offered by 
late-type eclipsing binaries to achieve a breakthrough in the reduction of the uncertainty
of the distance to the LMC, and to an understanding of the systematics affecting other methods
of distance determination.

\acknowledgments
WG, GP and DM gratefully acknowledge financial support for this
work from the Chilean Center for Astrophysics FONDAP 15010003, and from
the BASAL Centro de Astrofisica y Tecnologias Afines (CATA) PFB-06/2007. 
Support from the Polish grants N203 002 31/046, 
and N20303032/4275, and the FOCUS 
subsidy of the Fundation for Polish Science (FNP)
is also acknowledged. IBT acknowledges the support of NSF grant AST-0507325.
It is a pleasure to thank Willie Torres for sharing software with us, and 
to thank the support astronomers at ESO-La Silla and at Las Campanas Observatory
for their expert help in obtaining the observations. We also thank the ESO OPC and CNTAC
for allotting generous amounts of observing time to this project.
We dedicate this work to Bohdan Paczy{\'n}ski who encouraged us years ago
to exploit the route of eclipsing binaries towards an accurate distance
determination to the Magellanic Clouds.

\begin{figure}[p] 
\vspace*{8cm}
\includegraphics{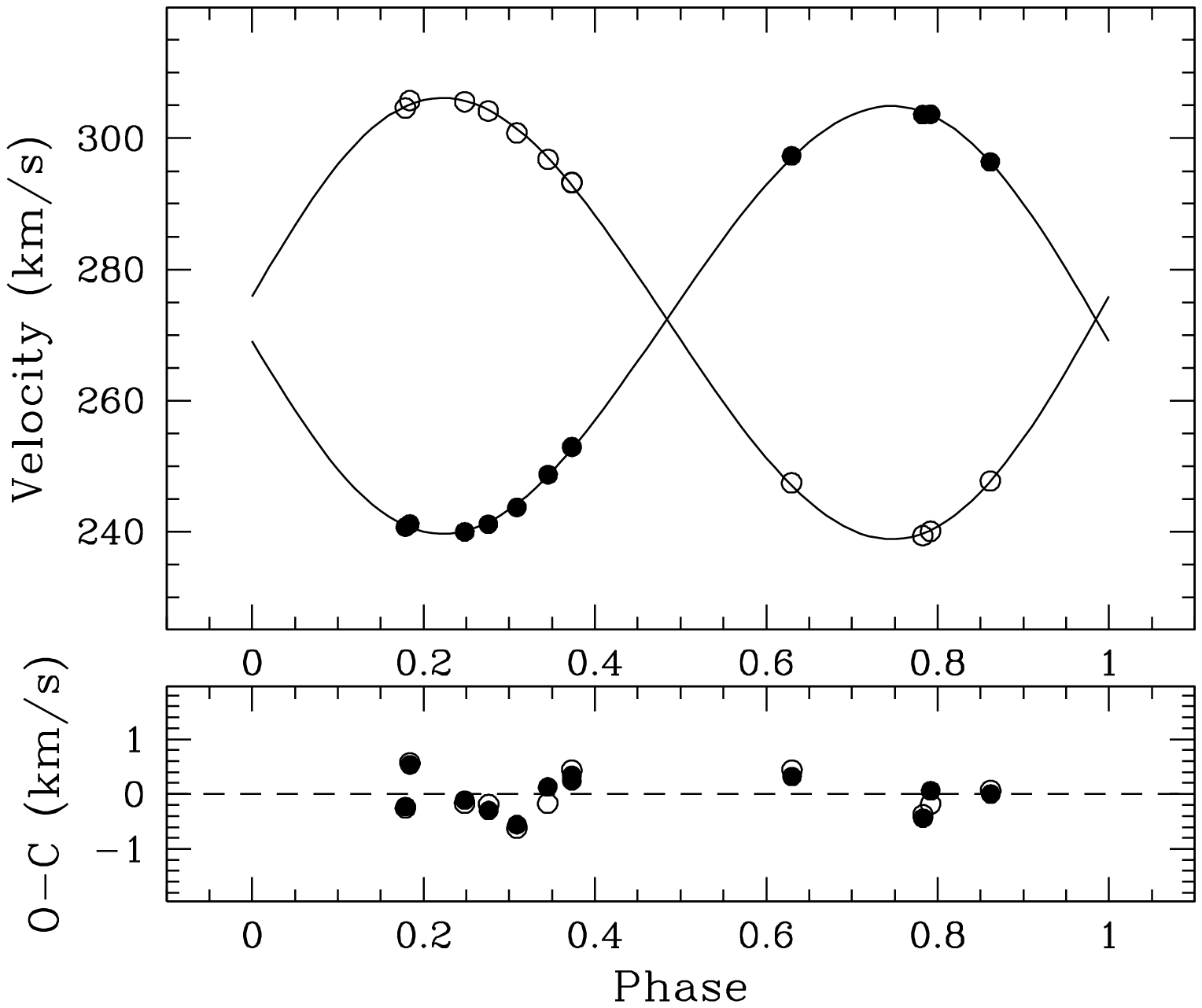} 
\caption{Spectroscopic orbit of OGLE-051019.64-685812.3.
}
\end{figure}  

\begin{figure}[p]
\vspace*{22 cm}
\includegraphics{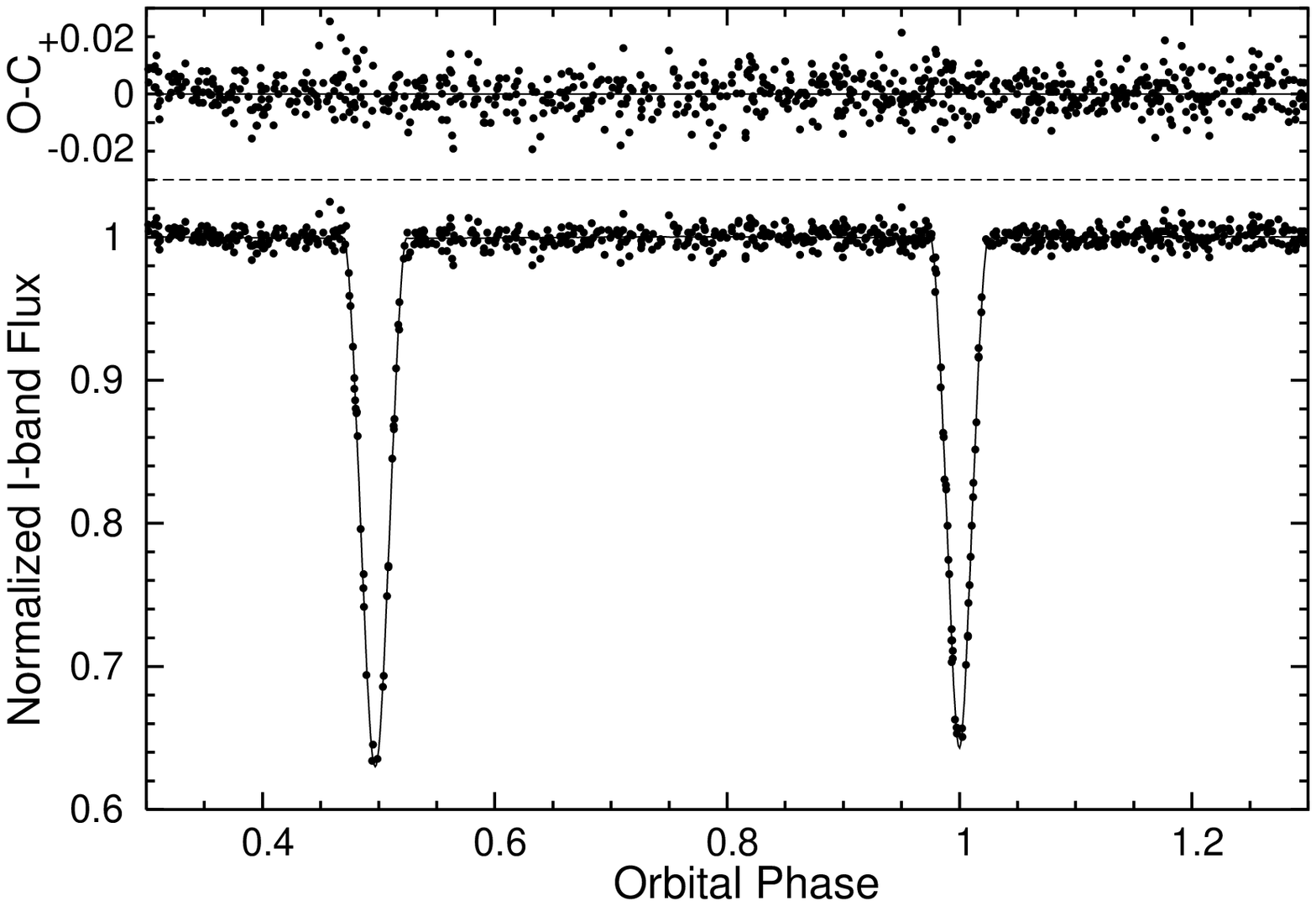}
\includegraphics{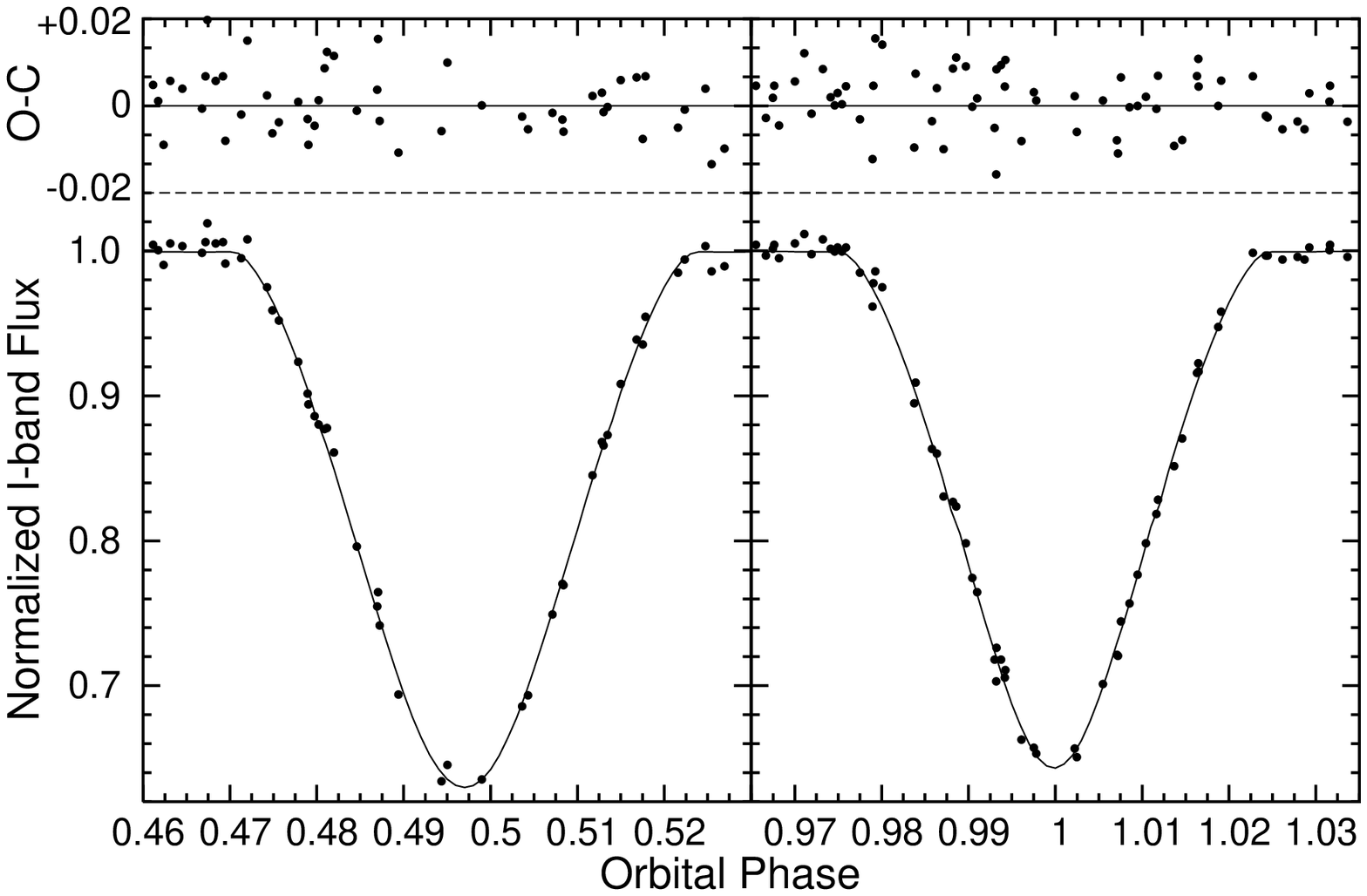}
\caption{Observed I band light curve together with the photometric solution,
as obtained from the analysis with the WD code. 
}
\end{figure}

%\begin{figure}[p]
%\vspace*{20 cm}
%\special{psfile=EB1838pr.ps  hoffset=-50  voffset=260 hscale=85 vscale=85}
%\special{psfile=EB1838se.ps  hoffset=-50  voffset=-40 hscale=85 vscale=85}
%\caption{Photometric solution
%}
%\end{figure}

\clearpage
\begin{deluxetable}{ccc}
\tablecaption{Radial Velocity Measurements for OGLE-051019.64-685812.3}
\tablehead{\colhead{HJD} & \colhead{$V_{1}$}
& \colhead{$V_{2}$} \\
\colhead{[days+2400000.0]} & \colhead{[km/s]}
& \colhead{[km/s]} }
\startdata
53989.90267 & 241.12 & 304.15\\
54004.84688 & 248.68 & 296.81\\
54010.79804 & 252.93 & 293.26\\
54010.84238 & 252.86 & 293.23\\
54065.79822 & 297.33 & 247.44\\
54098.63125 & 303.62 & 239.35\\
54183.52206 & 240.67 & 304.63\\
54184.60444 & 241.15 & 305.74\\
54314.91702 & 303.65 & 240.04\\
54329.88495 & 296.39 & 247.72\\
54412.73026 & 239.94 & 305.59\\
54425.78541 & 243.67 & 300.83\\
\enddata
\end{deluxetable}

\begin{deluxetable}{ccc}
\tablecaption{Effective temperature derived from the observed colors.
The mean value from all measurements ($T_{eff}$ = 5360 K) was adopted.}
\tablehead{\colhead{$T_{eff}$ [K]} & \colhead{calibration}
& \colhead{color} }
\startdata
5410 & Warthey and  Lee (2006) & J-K, V-K, V-I \\
5320 & Ramirez and Melendez (2005)      & V-I, V-K \\
5400 & Alonso et al. (1999)             & V-K \\
5300 & Di Bededetto  (1998)             & V-K \\
\enddata
\end{deluxetable}

\begin{deluxetable}{ccc}
\tablecaption{Astrophysical parameters of OGLE-051019.64-685812.3}
\tablehead{\colhead{} & \colhead{Primary} & \colhead{Secondary} }
\startdata
 P [days]         & 214.370 $\pm$ 0.008 & \\
 i [deg]          &     88.20 $\pm$ 0.10  & \\ 
 a [$R_{\bigodot}$] &     280.8 $\pm$ 1.1 & \\
 e                & 0.0395 $\pm$ 0.0025 & \\
 $\omega$  [deg]  & 96.53 $\pm$ 0.46 & \\
 $\varphi$        &  0.99850 $\pm$ 0.00003 \\
 $ q = m_{1} / m_{2}$ & 0.9695 $ \pm $ 0.0068 & \\ 
 $ \gamma$ [km/s]        &  272.39 $\pm 0.09 $ & \\
 K [km/s]         &  32.65 $\pm$ 0.14 & 33.67 $\pm$ 0.16 \\
$M/M_{\bigodot}$  &  3.29 $\pm$ 0.04   &  3.19 $\pm$ 0.04 \\
$R/R_{\bigodot}$  & 26.06 $\pm$ 0.28   & 19.76 $\pm$ 0.34 \\
$T_{eff} [K] $    &  5300 $\pm$ 100    & 5450 $\pm$ 100 \\
V [mag]           &  16.738  & 17.195 \\
I [mag]           &  15.969   & 16.466 \\
K [mag]           & 14.895   & 15.446 \\
distance [kpc]   & 50.4 $\pm 1.3$    & 50.0 $\pm 1.4$  \\
E(B-V)  [mag]    &  0.146 $\pm$ 0.02 & \\
 Fe/H & -0.5 dex (assumed) & \\
\enddata
\end{deluxetable}

\begin{deluxetable}{ccccc}
\tablecaption{Distance determinations to OGLE-051019.64-685812.3 based on different 
calibrations of the surface brightness color relations}
\tablehead{\colhead{d1 [kpc]} & d2 [kpc] & \colhead{luminosity class} &\colhead{color} & \colhead{reference} }
\startdata
51.4 & 50.7 & dwarfs & V-K & Di Benedetto (1998)\\
50.1 & 49.6 & giants & V-K & Di Benedetto (1998)\\
50.2 & 49.8 & dwarfs + giants & V-K & Di Benedetto (1998)\\
52.0 & 51.5 & dwarfs + giants & V-I & Di Benedetto (1998)\\
51.0 & 50.7 & dwarfs & V-K & Kervella et al.  (2004)\\
51.1 & 50.8 & dwarfs & V-K & Groenewegen (2004)\\
49.8 & 49.1 & giants & V-K & Groenewegen (2004)\\
50.4 & 50.0 & dwarfs + giants & V-K & Di Benedetto (2005)\\

\enddata
\end{deluxetable}

\end{document}